\documentclass[a4paper,11pt]{article}
\pdfoutput=1 

\usepackage{jheppub} 

\usepackage[T1]{fontenc} 

\newcommand{\mrm}[1]{\mbox{\rm #1}}
\newcommand{\be}{\begin{equation}}
\newcommand{\ee}{\end{equation}}
\newcommand{\br}{\begin{eqnarray}}
\newcommand{\bea}{\begin{eqnarray}}
\newcommand{\eea}{\end{eqnarray}}
\newcommand{\er}{\end{eqnarray}}
\newcommand{\ba}{\begin{array}}
\newcommand{\ea}{\end{array}}

\newcommand{\bi}{\begin{itemize}}
\newcommand{\ei}{\end{itemize}}
\newcommand{\bn}{\begin{enumerate}}
\newcommand{\en}{\end{enumerate}}
\newcommand{\bc}{\begin{center}}
\newcommand{\ec}{\end{center}}

\newcommand{\Eq}[1]{Eq.~(\ref{#1})}
\newcommand{\rfn}[1]{(\ref{#1})}

\newcommand{\beq}{\begin{equation}}
\newcommand{\eeq}{\end{equation}}

\newcommand{\abs}[1]{|#1|} 
\newcommand{\hc}[2][]{#2^{\dagger #1}} 
\newcommand{\T}[1]{#1^{T}}
\DeclareMathOperator{\tr}{tr}
\newcommand{\conj}[1]{#1^*}

\newcommand{\gsim}{\lower.7ex\hbox{$\;\stackrel{\textstyle>}{\sim}\;$}}
\newcommand{\lsim}{\lower.7ex\hbox{$\;\stackrel{\textstyle<}{\sim}\;$}}

\title{Embedding inflation into the Standard Model -- more evidence for classical scale invariance}

\author[a]{Kristjan Kannike}

\author[a]{Antonio Racioppi}

\author[b]{Martti Raidal}
\affiliation[a]{National Institute of Chemical Physics and Biophysics, R\"avala 10, 10143 Tallinn, Estonia}
\affiliation[b]{Institute of Physics, University of Tartu, Estonia}

\date{\today}

\emailAdd{antonio.racioppi@kbfi.ee}
\emailAdd{kristjan.kannike@cern.ch}
\emailAdd{martti.raidal@cern.ch}

\abstract{
If cosmological inflation is due to a slowly rolling single inflation field taking trans-Planckian values as suggested by the BICEP2 measurement of primordial tensor modes in CMB, embedding inflation into the Standard Model challenges standard paradigm of effective field theories. Together with an apparent absence of Planck scale contributions to the Higgs mass and to the cosmological constant, BICEP2 provides further experimental evidence for the absence of large $M_{\rm P}$ induced operators. We show that classical scale invariance -- the paradigm that all fundamental scales in Nature are induced by quantum effects -- solves the problem and
allows for a remarkably simple scale-free Standard Model extension with inflaton without extending the gauge group. Due to trans-Planckian inflaton values and vevs, a dynamically induced Coleman-Weinberg-type inflaton potential of the model can predict tensor-to-scalar ratio $r$ in a large range, converging around the prediction of chaotic $m^2\phi^2$ inflation for a large trans-Planckian value of the inflaton vev. Precise determination of $r$ in future experiments will single out a unique scale-free inflation potential, allowing to test
the proposed field-theoretic framework.
}

\begin{document}

\maketitle

\section{Introduction}

The Background Imaging of Cosmic Extragalactic Polarization (BICEP2) measurement of tensor modes from large angle Cosmic Microwave Background (CMB)
B-mode polarisation~\cite{Ade:2014xna}  confirms the last missing generic prediction of
the inflationary paradigm~\cite{Guth:1980zm,Linde:1981mu,Albrecht:1982wi} -- the existence of primordial tensor perturbations from gravity
waves~\cite{Mukhanov:1981xt,Bardeen:1983qw,Mukhanov:1990me}.  All collected experimental data from structure formation and CMB
anisotropies are consistent with the predictions of single field canonical slow-roll inflation.
In the present era of precision cosmology the main question has moved from finding another tree-level inflaton potential that predicts observables consistently with observations
(for reviews see~\cite{Olive:1989nu,Lyth:1998xn,2009arXiv0902.1529K}),
to the challenge of how to embed the inflation into the Standard Model (SM) consistently with our knowledge of particle physics and quantum field theory.

BICEP2 has detected primordial gravitational waves, measuring the tensor-to-scalar ratio to be~\cite{Ade:2014xna}
\bea
r=0.20^{+0.07}_{-0.05}.
\label{r}
\eea
After subtracting an estimated dust foreground the central value of the measurement may be reduced to $ r = 0.16^{+0.06}_{-0.05}.$
\Eq{r}  is in moderate tension with the upper bound $r<0.11$ given by the Planck Mission~\cite{Ade:2013uln}. We assume that this discrepancy
will be resolved by new data or by some other means~\cite{Mohanty:2014kwa}.
The BICEP2 result, if confirmed, corresponds to the Hubble parameter $H_*\sim 10^{14}$~GeV
and inflaton potential $V \sim (10^{16}~\mrm{GeV})^4$ during inflation. This implies, in a model-independent way via the Lyth bound~\cite{Lyth:1996im,Boubekeur:2005zm},
trans-Planckian values for the inflaton field. This also implies that the measured primordial curvature perturbations are dominantly generated by a
slowly rolling inflaton (see \cite{Okada:2014lxa} for an update of popular models), disfavouring curvaton scenarios~\cite{Lyth:2014yya} and allowing to rule in or rule out the slow-roll inflation with future measurements of
non-Gaussianity $f_{\rm{NL}}$~\cite{Lyth:2014yya,Kehagias:2014wza}.
Although alternative scenarios  can be saved by adding extra fields and dynamics to models,
in the following we shall assume that the BICEP2 result favours generic trans-Planckian single field slow-roll inflation.%
\footnote{If the change of $\epsilon$ is not monotonous, $r$ compatible with the BICEP2 results is possible for sub-Planckian inflation \cite{Choudhury:2013jya,Choudhury:2013iaa,Choudhury:2014kma}. In this case a more precise measurement of $\alpha_T$ wil decide between sub- and trans-Planckian inflation.}

Perhaps the most intriguing and most studied consequence of the BICEP2 result is the high scale of inflation.
Following the standard Wilsonian prescription, the inflaton $\phi$ potential can be written as
\bea
V=V_{\rm ren} + \sum_{n=5}^\infty c_n \frac{\phi^n}{M_{\rm P}^{n-4}},
\label{V}
\eea
where $V_{\rm ren}$ is the renormalisable part of inflaton potential, $M_{\rm P}=2.4\times 10^{18}$~GeV is the reduced Planck mass, and $c_n$ are
the Wilson coefficients of gravity-induced higher order operators. Since BICEP2 implies $\phi/M_{\rm P}\gsim (1\div 10)$~ \cite{Antusch:2014cpa}, the infinite sum of
non-renormalisable operators in \rfn{V} is badly divergent, predicting $V \gg (10^{16}~\mrm{GeV})^4$ and messing up
the inflation~\cite{Calmet:2014lga} and (meta)stability of the scalar potential \cite{Branchina:2013jra}.   At the same time, inflation models with a modified inflaton kinetic term,
 \bea
 L_{\rm kin}= \frac{1}{2} f\! \left( \frac{\phi}{M_{\rm P}}  \right) \partial_\mu \phi \partial^\mu \phi,
 \label{L}
 \eea
 where $f$ is a function of  $M_{\rm P}$-suppressed operators, have gained popularity after BICEP2. The reason is that after canonically normalising the kinetic term, the inflaton potential changes shape. This has been used to construct models consistent with \Eq{r}, for example to save Higgs inflation.%
 \footnote{The SM Higgs
 inflation~\cite{Bezrukov:2007ep} suffers from the vacuum stability problem~\cite{Degrassi:2012ry,Bezrukov:2012sa,Buttazzo:2013uya}, requires a non-minimal inflaton coupling to gravity, and predicts $r\ll 0.16.$
 To make Higgs inflation viable, vacuum must be made stable with help of
 extra singlet scalars~\cite{Kadastik:2011aa,Lebedev:2012zw,EliasMiro:2012ay,Gabrielli:2013hma,Ko:2014eia} or the top quark mass must be fine-tuned~\cite{Froggatt:1995rt,Isidori:2007vm} to a value that is
 disfavoured by experimental data by 2.2$\sigma$~\cite{Degrassi:2012ry} to 3.6$\sigma$~\cite{Kobakhidze:2014xda,Spencer-Smith:2014woa}, or extra fermions must be added~\cite{Haba:2014zda}.
 For recent papers on Higgs inflation scenarios see \cite{Masina:2014yga,Cook:2014dga,Hamada:2014iga,Nakayama:2014koa,Fairbairn:2014nxa,Bezrukov:2014bra,Enqvist:2014bua}.}
 Using non-renormalisable operators in the regime $\phi/M_{\rm P}>1$  is highly questionable.

 Trans-Planckian inflaton values have created a lot of confusion in physics community. The proposed solutions vary from assumptions that
 the unknown UV theory of gravity is such that all non-renormalisable operators are exponentially suppressed~\cite{Calmet:2014lga,Salvio:2014soa,Kaloper:2014zba,Chialva:2014rla}, to abandoning the inflation as the origin
 of density perturbations \cite{Kehagias:2014wza}.
However, in the light of successful experimental verification
 of all five generic predictions of the slow-roll inflation (almost scale-invariant density perturbations, adiabatic initial conditions, nearly Gaussian fluctuations,
 spatial flatness, and, finally, tensor perturbations from gravity waves) one should first study the implications of the BICEP2 result to our understanding of
 quantum field theories. This is the aim of our work.

 We argue that the apparent absence of the Planck scale induced operators \rfn{V} and \rfn{L}, as proven experimentally by the BICEP2 result,
 is an evidence for classically scale-free fundamental physics. This implies that all scales in physics are generated by quantum effects. We show that this
 paradigm can be extended also to inflation in phenomenologically successful way.  We present and embed a most minimal scale-free inflation model
 in the scale-free SM and show that the result is predictive and can be consistent with the BICEP2 and Planck measurements.

The idea of scale-invariant inflation is not new. Already the very first papers on inflation \cite{Linde:1981mu,Albrecht:1982wi,Linde:1982zj,Ellis:1982ws,Ellis:1982dg}
considered dynamically induced inflaton potentials {\it \`{a}~la} Coleman-Weinberg~\cite{Coleman:1973jx}.
 Since then  the Coleman-Weinberg inflation has been extensively studied in the context of grand unified theories
\cite{Langbein:1993ym,GonzalezDiaz:1986bu,Yokoyama:1998rw,Rehman:2008qs} and in $U(1)_{B-L}$ extension of the
SM~\cite{Barenboim:2013wra,Okada:2013vxa}. The common feature of all those models, probably inherited from the original Coleman-Weinberg paper~\cite{Coleman:1973jx},
is that the dynamics leading to dimensional transmutation is induced by new gauge interactions beyond the SM.
However, the dimensional transmutation does not need extra gauge interactions!
It can occur just due to running of some scalar quartic coupling $\lambda(\mu)\phi^4$ to negative values at some energy scale $\mu$ due to couplings to other scalar fields,
generating non-trivial physical potentials as demonstrated in Ref.~\cite{Heikinheimo:2013fta}. The models of this type are simple and generic, and therefore we
call them the minimal scale-free (or Coleman-Weinberg) models. In this paper we study this type of inflation models.

 Working consistently with the one loop effective inflaton potential that
is induced by inflaton couplings to other fields, we first study model-independent predictions of this scenario.
We find that the spectral index $n_s$ and the tensor-to-scalar ratio $r$ are strongly correlated in this scenario.
The latter can easily take large values measured by the BICEP2. In fact, the model predictions for $r$ converge
towards the prediction of chaotic inflation  potential $m^2\phi^2$ if the inflaton vacuum expectation value (vev) is
induced at trans-Planckian scales. This is the reason why our results differ from the results of Ref.~\cite{Barenboim:2013wra}.
Since our results are predictive, precise measurement of the tensor-to-scalar ratio in future experiments will essentially fix the inflaton potential
 of the minimal scale-free inflation scenario. After that we present the most minimal scale-free inflation model containing two
 real scalar fields in which this inflation scenario is realised, and work out model predictions explicitly.
We also work out constraints on the inflaton couplings to other fields in order not to spoil the predictions of inflation. We argue that the
 most natural way of reheating is the inflaton decays into the right-handed neutrinos consistently with successful leptogenesis.

 \section{Scale invariance and inflation}

Recently, the BICEP2 measurement of the tensor-to-scalar ratio $r$ is the second experimental result challenging fundamentals of the standard
 Wilsonian paradigm of renormalisation of quantum field theories. The first was the discovery of Higgs boson with mass $m_H=125.5$~GeV~\cite{Chatrchyan:2012ufa,Aad:2012tfa}.
 According to Wilsonian paradigm, marginal operators, such as the Higgs mass term $m^2|H|^2,$ should receive large, quadratically divergent
 contributions from any high scale they couple to. Therefore, no light scalars must exist in Nature and one anticipates $m_H^2 \sim M_{\rm P}^2.$
 This is nothing but the well known naturalness problem! The Higgs discovery, and the absence of supersymmetry  signal at the LHC,
 hint that the concept of naturalness must be revised~\cite{Heikinheimo:2013fta}.

 Similarly, the BICEP2 result implies the absence of non-renormalisable operators \rfn{V} and \rfn{L}
 that are suppressed by an explicit mass scale, $M_{\rm P}.$ By the common lore, all those operators must
 exist and be unsuppressed above $M_{\rm P}.$ To the contrary, the BICEP2 result shows us that particle physics seems to
 work above the Planck scale the same way it does below the Planck scale.

 Here we argue that there is a common solution to both problems -- the classical scale invariance of laws of physics (see \cite{Heikinheimo:2013fta,Hempfling:1996ht,Foot:2007as,Foot:2007ay,Foot:2007iy,Chang:2007ki,Holthausen:2009uc,Iso:2009ss,Iso:2009nw,Foot:2010av,Holthausen:2013ota,Steele:2013fka,Carone:2013wla,Dermisek:2013pta,Barenboim:2013wra,Hambye:2013dgv,Radovcic:2014rea,Englert:2013gz,Khoze:2013uia,Khoze:2013oga,Chun:2013soa,Hashimoto:2013hta,Khoze:2014xha,Guo:2014bha,Farzinnia:2014xia,Hashimoto:2014ela,Kubo:2014ova,Kubo:2014ida}).
 This means that the fundamental Lagrangian must contain no explicit mass scales, such as the electroweak
 scale Higgs mass or the Planck scale suppressed operators. All scales in physics, and we know that there are scales in physics,
 must be generated by dynamical dimensional transmutation. Here we accept that the BICEP2 result provides another experimental hint of the
 classical scale invariance as a fundamental law of nature.

Scale-free models of dynamical electroweak symmetry breaking, dark matter and inflation require just a minimal extension of the SM.
For example, in Ref.~\cite{Gabrielli:2013hma} all three tasks were achieved by extending the SM by only one complex singlet scalar field $S$
without imposing any additional symmetry for stabilising the dark matter. In this model the inflaton potential is of $\lambda_\phi \phi^4$ type (because $v_\phi \sim$ TeV) that
predicts large tensor-to-scalar ratio $r$ close to the $2\sigma$ upper limit of BICEP2 measurement.
This value is in clear contradiction with the Planck result. To lower the Ref.~\cite{Gabrielli:2013hma}
model prediction for $r$ to the experimentally preferred values $0.1\lsim r \lsim 0.2,$ the simplest solution is to introduce a non-minimal
coupling of inflaton $\phi=\mrm{Re}(S)$ to gravity \cite{Salopek:1988qh,Fakir:1990eg,Kaiser:1994vs,Komatsu:1999mt,Nozari:2007eq}, $\xi \phi^2 R,$ where $R$ is the Ricci scalar. Such a coupling flattens the inflaton potential at
high scales and the model predictions for $r$ can be tuned to the required range. We leave studies of such a inflation model to a future publication~\cite{future}.
In this work we choose a different approach -- we neglect the non-minimal coupling, add the inflaton $\phi$ as a new field and assume that all measured inflation observables
are induced by a scale-free scalar theory. We show that scale-free theory can predict the main inflation observables, $(n_s,\,r)$, in a narrow correlated range.
Future precision measurements of these observables will essentially fix the model parameters.

\section{Properties of single field scale-free inflation scenario}
We start by studying generic properties of single field scale-free inflation scenario.
First we take a model-independent approach and assume that the shape of the potential is
generated dynamically by one-loop effects without specifying the underlying physics.
Therefore our approach can be viewed as an effective Lagrangian one.
One concrete model realisation of this type of potentials will be presented in the next section.

Our approach in this section is essentially the same as in Ref.~\cite{Barenboim:2013wra} except that we consider both the
small and the large field inflation scenarios. However, our results and, therefore, conclusions differ substantially. We shall
explain this difference in results shortly.

The tree-level Lagrangian to start with is
\be
 V =\Lambda^4  + \frac{\lambda_\phi}{4} \phi^4  ,
 \label{V0}
\ee
where $\phi$ is the inflaton field and $\Lambda$ is the cosmological constant needed to tune
the potential of the minimum to zero. While particle physics observables depend only on the difference of
the potential, gravity couples to the absolute scale creating the cosmological constant problem that so far does not have a commonly accepted elegant solution. In the following we view the existence of $\Lambda$
as a phenomenological necessity and accept the fine tuning associated with it.

\begin{figure}[t!]
\centering
 \includegraphics[width=0.7\textwidth]{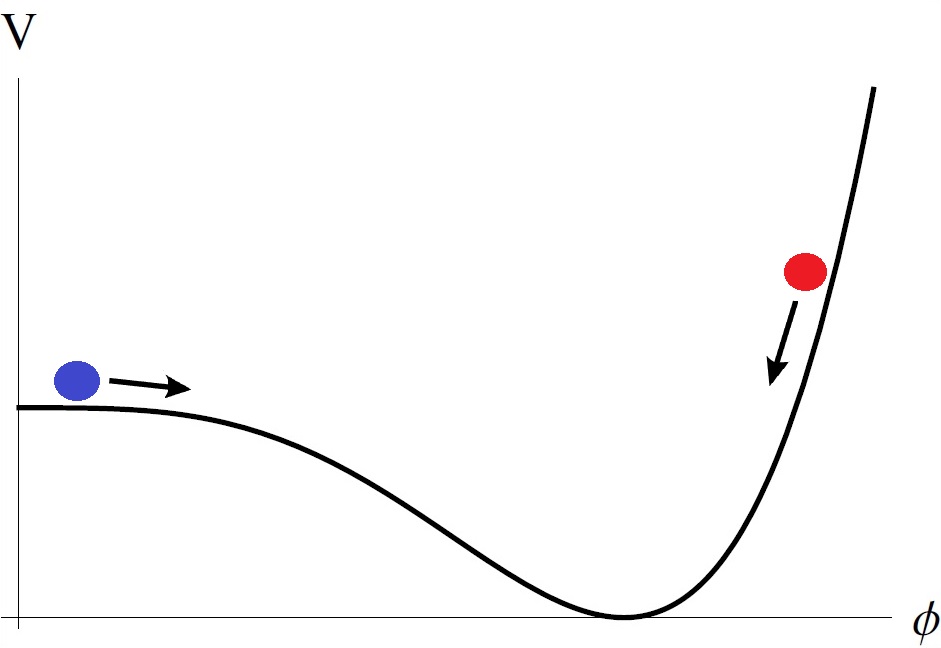}
 \caption{The shape of scale-free inflaton potential. Both chaotic (red) and hilltop (blue) inflation are allowed.}
 \label{Vfig}
\end{figure}

In realistic models of inflation one has to consider effects of inflaton couplings to itself and to other fields.
Working consistently at one loop level, one obtains from  \rfn{V0}  the renormalisation group improved effective inflaton potential
 {\it \`{a}~la} Coleman-Weinberg~\cite{Coleman:1973jx},
\be
 V_{\rm eff} = \Lambda^4 + \frac{\beta_{\lambda_\phi}}{4} \ln \left|\frac{\phi}{\phi_0} \right| \phi^4
 \label{Vrunbeta},
\ee
where the beta-function $\beta_{\lambda_\phi}$ describes running of $\lambda_\phi$ due to inflaton couplings and $\phi_0$ is the
scale induced by dimensional transmutation that is closely related to the minimum of the potential.
Unlike the previous works
\cite{Linde:1981mu,Albrecht:1982wi,Langbein:1993ym,GonzalezDiaz:1986bu,Yokoyama:1998rw,Rehman:2008qs,Barenboim:2013wra}, we will make do \emph{without} extending the gauge symmetries of the SM.
Computational details will be presented in the next section.
The shape of the potential is illustrated in Fig.~\ref{Vfig}. This is an example how a scale of physics is generated
in initially scale-invariant theory.

For further use it is convenient to rewrite the potential \rfn{Vrunbeta} as
\be
 V_{\rm eff} = \Lambda^4 \left(1+ \frac{\beta_{\lambda_\phi}}{4 \Lambda^{4}} \ln \left|\frac{\phi}{\phi_0}  \right| \phi^4 \right) ,
 \label{V2}
\ee
where we have factorised out the constant $\Lambda$.
This potential has a minimum at
 \be
  v_\phi = \frac{\phi_0 }{\sqrt[4]{e}} .
  \label{vevphi}
 \ee
According to our assumptions,  the cosmological constant $\Lambda$ is adjusted so that $V(v_\phi)=0$.
This is needed in order to avoid $\phi=v_\phi$ as an allowed inflaton initial configuration that leads to eternal inflation and the concerning issues \cite{Guth:2007ng}.
Solving this constraint for $\Lambda$, we get
\be
 \Lambda = \phi_0 \sqrt[4]{\frac{\beta_{\lambda_\phi}}{16 e}}.
 \label{lam}
\ee
We do not attempt to solve the cosmological constant problem in this work and consider \Eq{lam} as a phenomenological necessity.

The potential \rfn{V2} represents a dynamical realization of the inflaton potential.
As depicted in Fig.~\ref{Vfig}, such a shape allows for two different, generic types of inflation depending on the initial conditions:
\begin{itemize}
 \item[i.] Small-field hilltop inflation, when $\phi$ rolls down from small field values towards $v_\phi$
 \item[ii.] Large-field chaotic inflation, when $\phi$ rolls down from large field values  towards $v_\phi$.
\end{itemize}
Following~\cite{2009arXiv0902.1529K},  it is straightforward to compute the slow roll parameters, number of $e$-folds $N$, spectral index $n_s$ and its scale dependence,
tensor-to-scalar ratio $r$ and other inflation observables for the potential \rfn{V2}. We study for which parameter space this potential
can support phenomenologically acceptable inflation.

\begin{figure}[t!]
\centering
 \includegraphics[width=0.7\textwidth]{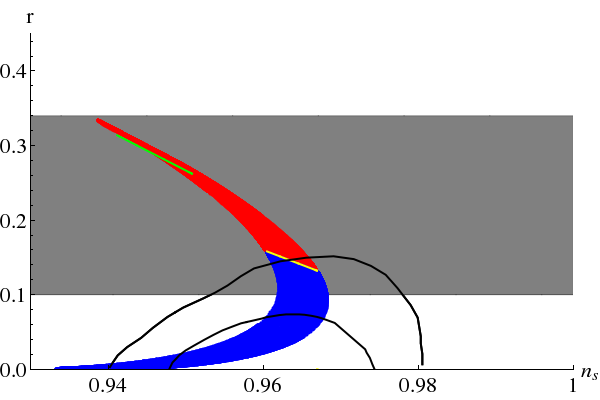}
 \caption{Predictions for tensor-to-scalar ratio $r$ as a function of $n_s$ for $N \in [50,60]$ $e$-folds.
  The blue region represent the hilltop inflation configuration while the red one the chaotic inflation.
  For reference we also plot predictions of  $\frac{m^2}{2} \phi^2$ (yellow) and $ \l \phi^4$ (green) potentials.
The $2\sigma$ BICEP2 band and $2\sigma,$ $1\sigma$ Planck bounds are also presented.
}
 \label{rvsn}
\end{figure}

The result in the $(n_s,r)$ plane is presented in Fig.~\ref{rvsn}.
The red and the blue regions correspond to the predictions of our model producing $N \in [50,60]$ $e$-folds of inflation.
We considered $\phi_0$ in the range\footnote{This is a crucial difference with \cite{Barenboim:2013wra}, in which they consider $v_\phi < M_{\rm P}$,
which automatically implies that the standard paradigm of inflation cannot work in the hilltop region, since one cannot satisfy the
Lyth bound~\cite{Lyth:1996im,Boubekeur:2005zm}.}
$[0.1,1000] \, M_{\rm P}$.

The blue region represent the hilltop inflation configuration while the red one corresponds to the chaotic inflation.
For reference and for interpretation of our result we also plot the predictions of
$V =  \frac{m^2}{2} \phi^2$ (yellow) and    $V =  \l \phi^4$ (green) potentials.
The hilltop inflation takes place in the region under the yellow line, while chaotic inflation occurs in the region above  it.
The chaotic region, of course, also contains the simple $\l \phi^4$ model.
The  grey band represents the $2\sigma$ BICEP2 result while the black lines are the $1\sigma$ and $2\sigma$ Planck bounds.

\begin{figure}[t!]
\centering
 \includegraphics[width=0.7\textwidth]{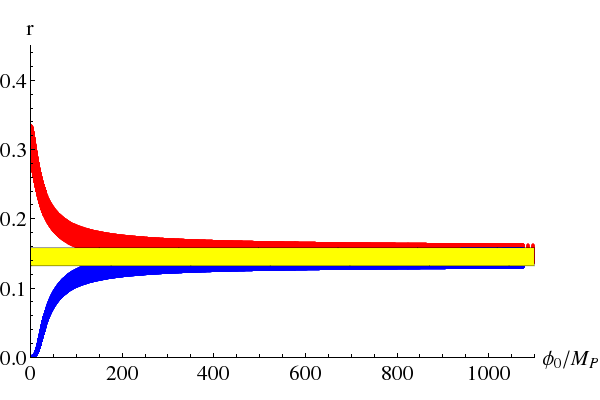}
 \caption{The predicted range for $r$ in function of $\phi_0$  producing $N \in [50,60]$ $e$-folds of inflation in our model and in the $\frac{m^2}{2} \phi^2$ inflation. The colour code is the same as in previous figures. }
 \label{rvsfi}
\end{figure}

It follows from Fig.~\ref{rvsn} that the spectral index and the tensor-to-scalar ratio are strongly correlated in the considered scale-free inflation scenario.
Present experimental accuracy allows for quite large model parameter space
consistent with the BICEP2 result, with the Planck result, and with the both. In particular, with the present accuracy
the data cannot not distinguish between the hilltop and the chaotic inflation.
With more data, however, this situation may change and experimental data would be able to uniquely identify the inflaton potential of our
model. After BICEP2, the $\frac{m^2}{2} \phi^2$ chaotic inflation potential has got a lot of attention since its predictions
agree well with experimental results. A particularly interesting conclusion of our work is that the scale-free inflation with non-vanishing
inflaton couplings can reproduce the same result. In our case this corresponds to the limit of very large inflaton vev, $v_\phi \gg M_{\rm P}.$
In this limit the shape of inflaton potential around the minimum becomes symmetric and inflation observables
lose sensitivity to the initial conditions. To explain this limit better, we plot in Fig. \ref{rvsfi} the predicted range for $r$ in function of $\phi_0$
producing $N \in [50,60]$ $e$-folds of inflation in our model and in the $\frac{m^2}{2} \phi^2$ inflation. The colour code is the same as in previous figures.
We can see that for $v_\phi \gg M_{\rm P}$, the three different regions overlap.
Therefore, if future data will determine  $(n_s,r)$ along the yellow line in Fig.~\ref{rvsn}, this will
support the scale-free inflation with a trans-Planckian inflaton vev.

We note that the parameter space considered in Ref.~\cite{Barenboim:2013wra} corresponds to the tail of blue region in Fig.~\ref{rvsn}
with $r\ll 0.1$ and $n_s< 0.945.$ Therefore those authors mistakenly concluded  that the scale-free inflation is not consistent with Planck results.
The reason for that is that they considered only inflaton field values below the Planck scale. In our opinion this assumption is overly restrictive.

\begin{figure}[t!]
\centering
 \includegraphics[width=0.7\textwidth]{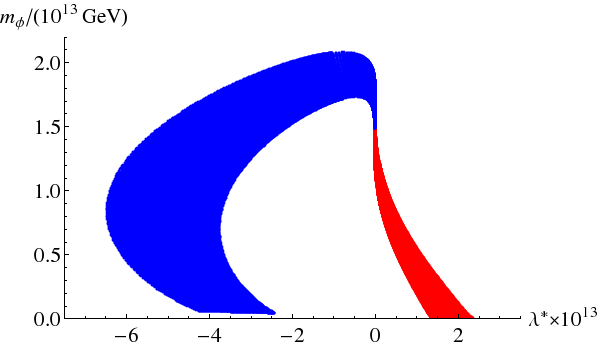}
 \caption{The range for $m_\phi$ and $\lambda^*$ corresponding to the results in Fig.~\ref{rvsn}.  The colour code is the same as in previous figures. }
 \label{mfivslambda}
\end{figure}

To explain our results in more detail we present in Fig.~\ref{mfivslambda} dynamically generated values for the inflaton mass $m_\phi$ against the
inflaton quartic self-coupling $\lambda_\phi^*$ at the beginning of inflation (denoted by the star). The colour code is the same as in Fig.~\ref{rvsn}.
We see that the phenomenological requirement is that the inflaton mass is always of the order ${\cal O}(10^{13})$~GeV.
We also see that the hilltop inflation, of course, starts only when $\lambda_\phi^*$ is negative,
while the chaotic inflation, instead, can take place for positive $\lambda_\phi^*$. This is the result of dimensional transmutation.
The absolute magnitude of $\lambda_\phi^*$ is tiny, ${\cal O}(10^{-13})$.
In scale-free inflation scenarios the existence of small scalar couplings have natural phenomenological explanation.
Namely, dimensional transmutation occurs when $\lambda_\phi$ runs through zero value. Thus the inflation naturally
occurs when $ \lambda_\phi^*\approx 0$ at very high scales.

As expected in slow roll models, the running of the spectral index  $\frac{d n_s}{d \ln k}$ in this framework is quite small, $\frac{d n_s}{d \ln k} \in [-0.0014,0.0045]$.
In particular, $\frac{d n_s}{d \ln k} \in [0.0005,0.0045]$ for the chaotic inflation scenario and $\frac{d n_s}{d \ln k} \in [-0.0014,0.0008]$ for the hilltop inflation scenario.

\section{The minimal scale-free model for inflation}
The results presented in previous section are model independent.
In this section we present the minimal scale-free inflation model giving rise to the inflaton potential \rfn{V2}.
As explained before, by minimal we mean no new gauge interactions beyond the SM, the dimensional transmutation is induced by singlet inflaton
coupling with another singlet.

We consider the following Lagrangian that extends the SM by two real singlet scalar field $\phi$ and $\eta$ and three heavy singlet right-handed neutrinos $N_{i}$:
\begin{align}
 {\cal L} &= {\cal L}_{\rm SM} + \frac{1}{2}\partial_\mu\phi \partial^\mu\phi +  \frac{1}{2}\partial_\mu\eta \partial^\mu\eta + {\cal L}_Y - V, \\
  {\cal L}_Y &= Y_{N}^{ij} \bar{L}_{i} i \sigma_{2} H^{*} N_{j} + {\rm h.c.} + Y_\phi^{ij} \bar N^c_{i }N_{j} \phi \nonumber \\
  & +  Y_\eta^{ij} \bar N^c_{i }N_{j} \eta,  \\
 V& = \Lambda^4 + \frac{1}{2} \lambda_{h\phi} \abs{H}^{2} \phi^{2} + \frac{1}{2} \lambda_{h\eta} \abs{H}^{2} \eta^{2}  \label{V3} \nonumber \\
 & + \frac{\lambda_\phi}{4} \phi^4 + \frac{\lambda_{\phi \eta}}{4} \eta^2 \phi^2 + \frac{\lambda_\eta}{4} \eta^4,
\end{align}
where $ {\cal L}_Y$ presents scalar Yukawa couplings with the right-handed neutrinos needed to generate small neutrino masses via seesaw
mechanism~\cite{Minkowski:1977sc,Yanagida,GellMann:1980vs,Glashow:1979nm,Mohapatra:1979ia,Schechter:1980gr,Schechter:1981cv}, leptogenesis~\cite{Fukugita:1986hr}
and reheating of the Universe. $H$ is the SM Higgs field and $L$ is the left-handed lepton doublet.
We assume that the kinetic terms are canonically normalised and there are no explicit mass terms in the scalar potential $V$.
The couplings in  ${\cal L}_Y$ run according to the coupled set of renormalisation group equations (RGEs)
given in the Appendix \ref{app:rges}. We derived the RGEs using the PyR@TE package~\cite{Lyonnet:2013dna}.

The one-loop inflaton effective potential can be written as
\begin{equation}
V_{\rm eff }= V + \Delta V,
\end{equation}
where the loop-level contribution reads
\begin{equation}
\begin{split}
\Delta V &= \frac{1}{64 \pi^2} \left[ \sum_{i=1}^2 m_i^4 \left(\ln \frac{m_i^2}{\mu^2} - \frac{3}{2} \right) \right. \\
&\left. - 2 \tr{\left \{ M_{N} \hc{M_{N}} \left(\ln \frac{M_{N} \hc{M_{N}}}{\mu^2} - \frac{3}{2} \right) \right\} } \right].
\end{split}
\end{equation}
Here $m_{i}^2$ are the eigenvalues of the field-dependent scalar mass matrices
\begin{equation}
m_{\phi\eta}^2 =
\begin{pmatrix}
3 \lambda_\phi \phi^2 + \frac{1}{2} \lambda_{\phi\eta} \eta^2 & \lambda_{\phi\eta} \phi \eta \\
 \lambda_{\phi\eta} \phi \eta & 3 \lambda_\eta \eta^2 + \frac{1}{2} \lambda_{\phi\eta} \phi^2
\end{pmatrix},
\end{equation}
\begin{equation}
M_N = Y_\phi \phi + Y_\eta \eta,
\end{equation}
and $\mu$ is the renormalisation scale. We have neglected the Higgs contributions to $\Delta V$ since those are proportional to the portal
couplings $\lambda_{h\phi}$ and $\lambda_{h\eta}$ that are constrained to be negligibly small due to their contributions to the Higgs boson mass
(see numerical estimates below).
The Higgs field will act as light spectator without giving sizeable contributions to inflationary or post-inflactionary dynamics.
For cases in which the Higgs field has a more active role, see \cite{Enqvist:2014bua,Enqvist:2013kaa,DeSimone:2012qr,Lyth:2005qk,Dvali:2003em,Kofman:2003nx}.
Inflation will take place in the direction $\eta=0$, which is the
minimal value  for the field $\eta$. We will see later that such an
assumption is self-consistent. The RGE improved effective
potential for the direction of $\phi$ reads
\begin{align}
V_{\rm eff} &=\Lambda ^4 + \frac{ \lambda _{\phi } (\mu) \phi^4}{4}
     +\frac{  \lambda _{\phi \eta
   }^2} {256 \pi ^2}    \left(\ln \frac{\phi ^2 \lambda _{\phi \eta }}{2 \mu
   ^2}-\frac{3}{2}\right)    \phi ^4, \nonumber
\end{align}
where we neglect the $\lambda_\phi$ contribution (since it must be extremely small, see Fig.~\ref{mfivslambda})
and the heavy neutrino contributions at one loop level. The beta function $\beta_{\lambda_\phi}$ is dominated by $\lambda_{\phi \eta}$.
In the previous section we treated $\beta_{\lambda_\phi}$ as a constant. It can be easily checked that also this assumption is consistent.
 Therefore
\be
 V_{\rm eff} =\Lambda ^4 + \frac{ \lambda _{\phi \eta }^2 }{{256 \pi ^2}}
 \left(\ln \frac{\mu^{2} }{\mu _0^{2}} + \ln \frac{ \lambda _{\phi \eta } \phi^2}{2 \mu ^2}-\frac{3}{2} \right) \phi^4,
\ee
where $\mu_{0}$ is defined by $\lambda_{\phi}(\mu_{0}) = 0$.
We can eliminate the dependence on the renormalisation scale, getting
\be
 V_{\rm eff}=\Lambda ^4 + \frac{ \lambda _{\phi \eta }^2}{256 \pi ^2} \left[ \ln \left(\frac{
   \lambda _{\phi \eta } \phi ^2}{\mu _0^2}\right)-\frac{3}{2} \right] \phi^{4} .
\ee
It can be easily checked that $\lambda_{\phi \eta}$ should be small enough to
be approximated  $\beta_{\lambda_{\phi \eta}} \sim 0$, therefore  $\lambda_{\phi \eta}$ can be treated as a constant value and so $\beta_{\lambda_\phi}$.

Then we can make a simple reparametrization getting
\be
  V_{\rm eff} =\Lambda ^4+\frac{\lambda_{\phi \eta} ^2  \ln \frac{\phi^{2}
   }{\phi_0^{2}}}{256 \pi ^2} \phi ^4,
\ee
where $\phi_0=\sqrt{\frac{2 e^{3/2}}{\lambda _{\phi \eta }}} \mu _0 $. We
see that we have reproduced the potential Eq.~(\ref{Vrunbeta}).
Therefore, the results presented in the previous
section all hold for this model, up to a proper redefinition of the
parameters. 

\begin{figure}[t]
\centering
 \includegraphics[width=0.7\textwidth]{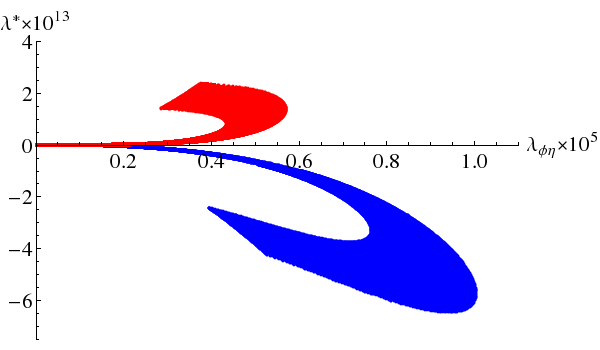}
 \caption{The range for $\lambda^*$ and $\lambda_{\phi\eta}$. The colour code is the same as in previous figures. }
 \label{portalvslambda}
\end{figure}

In Figure \ref{portalvslambda} we give the plot $\lambda^*$ in function of the portal coupling $\lambda_{\phi\eta}$. The colour code is the same as in previous figures. We can appreciate again that  $\lambda^*$ is positive for large field inflation, while it is negative for the hilltop one. $\lambda^*$ approaches zero in the region in which the potential gets closer to the $m^2 \phi^2$ potential. We see that $\lambda_{\phi\eta} \lesssim 10^{-5}$.
Moreover, the scalar $\eta$ gets a mass via portal coupling
\footnote{In a similar way we could also generate dynamically the Higgs boson mass term, via the portal $\lambda_{h \phi} \phi^2 h^2$.
However this will require an extremely fine tuned $\lambda_{h \phi}$, since $v_\phi > M_{\rm P}$. It is not in the purpose of this paper to discuss how the electroweak scale is generated and how to consistently embed inflation and SM in a full classical scale invariant model. We leave such studies to a future publication~\cite{future}.} as
 \be
  m_\eta = \sqrt{\frac{\lambda_{\phi \eta}}{2}} v_\phi.
 \ee
For the presented inflation scenario to be consistent, the self-coupling of $\eta$ has to be positive, $\lambda_{\eta} > 0$, but otherwise can have any value.
We remind the reader that in this scenario $v_\phi$ tends to be very large, typically above Planck scale.
Therefore very large particle mass $ m_\eta$ can be obtained for relatively small $\lambda_{\phi \eta}.$

\begin{figure}[t]
\centering
 \includegraphics[width=0.7\textwidth]{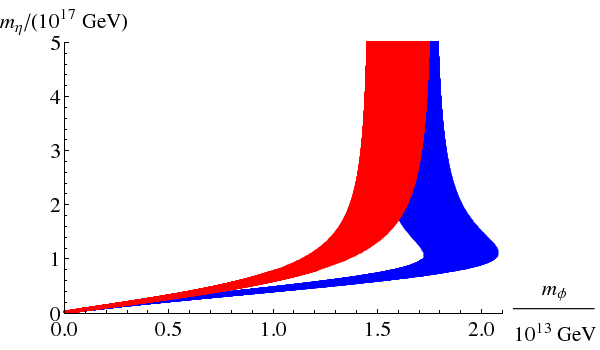}
 \caption{The range for $m_\eta$ and $m_\phi$. The colour code is the same as in previous figures. }
 \label{metavsmfifinal}
\end{figure}

In Fig.~\ref{metavsmfifinal}  we present the range for $m_\eta$ and $m_\phi$. As before, the
blue region represent the hilltop inflation configuration while
the red one the chaotic inflation. We see that $m_\eta \sim 10^{17}$~GeV
which is one order of magnitude larger than the inflation scale
$\sim 10^{16}$~GeV, and several orders of magnitude larger than $m_\phi$.
Therefore, it is natural to assume that $\eta$ is decoupled  and is
frozen at its minimum $\eta=0$ during inflation. Otherwise the $\eta$ potential will dominate
the energy density and inflation would not occur. It has been shown in Ref.~\cite{Lebedev:2012sy}
that $\eta=0$ is quickly achieved during inflation due to the portal coupling $\lambda_{\phi \eta}.$
Thus the model is self-consistent.

So far we have shown that the inflaton couplings may be responsible for generating the inflaton potential at one loop.
We also know that inflaton must decay and reheat the Universe. Let us consider constraints on inflaton couplings from reheating.
The reheating temperature induced by the inflaton decays is given by
\be
 T^{\rm RH} = \left( \frac{90}{\pi^2 g_*} \right)^{\frac{1}{4}} \sqrt{\Gamma M_{\rm P}},
\ee
where $\Gamma$ is the total width of the inflaton, and  $g_*$ is the number of relativistic degrees of freedom in the model.
In our scenario reheating cannot happen via inflaton decays into the Higgs boson pairs because the Higgs-inflaton coupling $\lambda_{h \phi} \phi^2 h^2$ is constrained to be extremely small due to very large inflaton vev ($v_\phi > M_{\rm P}$) contribution to the Higgs mass. Supposing that the portal gives a tree level contribution to the Higgs mass of order  1~GeV, which is around $3 \sigma$ uncertainty of the present Higgs mass, for $v_\phi \simeq 1.4 \times 10^{20}$~GeV we get  $\lambda_{h \phi} \simeq 5 \times 10^{-41}$ and a reheating temperature $T^{\rm RH}_H \sim 7 \times 10^{-20}$ GeV, which is phenomenologically unacceptable.

Instead, reheating and leptogenesis can take place via inflaton decays directly into the right-handed neutrinos driven by the Yukawa couplings $Y_\phi$.
Such Yukawas are constrained by the requirement that they would not spoil the inflation potential,
and that the right-handed neutrino masses are smaller than at least a half of the inflaton mass.
It can be easily checked that the most constraining one is the last bound, which implies $Y_\phi \lesssim 0.1 \lambda_{\phi \eta}$, which ensures that the Yukawa loop contribution to  $V_{\rm eff}$ are negligible. For example, for $\Lambda \simeq 2.5 \times 10^{16}$~GeV and $v_\phi \simeq 1.4 \times 10^{20}$~GeV, we get $r \simeq 0.1$.
If the inflaton does not have any other sizeable coupling but $Y_\phi$, as we assume here, it directly decays to $N_i$ which decay asymmetrically to $\bar LH$ and $L\bar H$
generating the baryon asymmetry and reheating the Universe. For details of this scenario see Ref.~\cite{Giudice:2003jh}.

\section{Discussion and Conclusions}

The BICEP2 measurement \rfn{r}, that we assume will be confirmed by other experiments, motivated us to study predictions of classically scale-free single field inflation.
In this scenario the inflaton potential as well as all mass scales are generated dynamically
via dimensional transmutation due to inflaton couplings to other fields.
This approach is conceptually different from a typical inflation study which considers a tree-level inflaton potential and
checks that all other effects are small. Therefore, to achieve our aims, we have worked consistently at one loop and computed
the one-loop effective inflaton potential improved with one-loop renormalisation group equations presented in Appendix A.
 Since the inflaton field must take trans-Planckian values, gravity above the Planck scale must couple weakly to particle physics.
 Classical scale invariance provides also a natural solution to the absence of large Planck suppressed non-renormalisable operators.
Such classically scale-free models of gravity have been proposed recently in Ref.~\cite{Salvio:2014soa}.

First, working model independently with Coleman-Weinberg type single inflaton potential \rfn{V2}, we computed which values of $r$ and $n_s$
the model can accommodate. The results show that $r$ and $n_s$ are strongly correlated but the present experimental accuracy does not allow to specify the model parameters,
and almost any value of $r$ is, in principle, achievable. However, if future measurements will determine $r$ and $n_s$ with high accuracy, this
scenario must pass non-trivial tests. Interestingly, if the future result is consistent with the prediction of  tree level potential $m^2\phi^2,$
in the scale-free inflaton scenario this corresponds to very large scale inflaton physics with its vev above the Planck scale.

We have presented a minimal scale-free inflaton model and shown with explicit computation how the inflaton potential arises
from dimensional transmutation. We computed RGEs for this model and present those in the Appendix~\ref{app:rges}. We conclude that
classically scale-free inflation models are attractive, self-consistent framework to address physics above Planck scale.

\acknowledgments

The authors thank Oleg Lebedev for useful discussions and Florian Lyonnet for kind assistance with the PyR@TE package.
This work was supported by  grants MJD298,  MTT60,  IUT23-6, CERN+, and by EU through the ERDF  CoE program.

\clearpage

\appendix

\section{One-Loop RGEs}
\label{app:rges}

The one-loop RGEs of the scalar quartic couplings and the Yukawa couplings are given by

\begin{align}
16 \pi^2 \beta_{\lambda_h}
& = \beta_{\lambda_h}^{\rm SM} + \frac{1}{2} (\lambda_{{h\phi}}^{2} +\lambda_{{h\eta}}^{2})
+ 2 \lambda_{h} ( \tr{Y_{N}^{\dag}Y_{N} } + \tr{\conj{Y_{N}}\T{Y_{N} } } ) - \tr{Y_{N}Y_{N}^{\dag}Y_{N}Y_{N}^{\dag} } \\
& - \tr{\T{Y_{N} }\conj{Y_{N}}\T{Y_{N} }\conj{Y_{N}} }, \notag \\
16 \pi^{2} \beta_{\lambda_{h\phi}} &= 4 \lambda_{{h\phi}}^{2} +\lambda_{{\phi\eta}} \lambda_{{h\eta}} + 6 \lambda_{{h\phi}} (2 \lambda_{h} + \lambda_{\phi})
- \lambda_{{h\phi}} \left( \frac{3}{2}  g^{2} + \frac{9}{2} g^{\prime 2} \right)
+ \lambda_{{h\phi}} [\tr{Y_{{e}}^{\dag}Y_{{e}} } \\
& + \tr{\conj{Y_{{e}}}\T{Y_{{e}} } } +3 \tr{Y_{{d}}^{\dag}Y_{{d}} } + 3  \tr{\conj{Y_{{d}}}\T{Y_{{d}} } }  + 3  \tr{Y_{{u}}^{\dag}Y_{{u}} }
+ 3  \tr{\conj{Y_{{u}}}\T{Y_{{u}} } } + \tr{\conj{Y_{N}}\T{Y_{N} } } \notag \\
& +  \tr{Y_{N}^{\dag}Y_{N} } \notag  +8  \tr{Y_{{\phi}}^{\dag}Y_{{\phi}} } ] \notag - 8 \tr{Y_{{\phi}}Y_{N}^{\dag}Y_{N}Y_{{\phi}}^{\dag} }
- 8 \tr{Y_{{\phi}}Y_{{\phi}}^{\dag}\T{Y_{N} }\conj{Y_{N}} }
- 8 \tr{Y_{N}Y_{{\phi}}^{\dag}Y_{{\phi}}Y_{N}^{\dag} } \notag \\
&- 8 \tr{\T{Y_{N} }\conj{Y_{N}}Y_{{\phi}}Y_{{\phi}}^{\dag} }, \notag \\
16 \pi^{2} \beta_{\lambda_{h\eta}} &= 4 \lambda_{{h\eta}}^{2} +\lambda_{{\eta\eta}} \lambda_{{h\eta}} + 6 \lambda_{{h\eta}} (2 \lambda_{h} + \lambda_{\eta})
- \lambda_{{h\eta}} \left( \frac{3}{2}  g^{2} + \frac{9}{2} g^{\prime 2} \right)
+ \lambda_{{h\eta}} [\tr{Y_{{e}}^{\dag}Y_{{e}} } \\
& + \tr{\conj{Y_{{e}}}\T{Y_{{e}} } } +3 \tr{Y_{{d}}^{\dag}Y_{{d}} } + 3  \tr{\conj{Y_{{d}}}\T{Y_{{d}} } }  + 3  \tr{Y_{{u}}^{\dag}Y_{{u}} }
+ 3  \tr{\conj{Y_{{u}}}\T{Y_{{u}} } } + \tr{\conj{Y_{N}}\T{Y_{N} } } \notag \\
& +  \tr{Y_{N}^{\dag}Y_{N} } \notag  +8  \tr{Y_{{\eta}}^{\dag}Y_{{\eta}} } ] \notag - 8 \tr{Y_{{\eta}}Y_{N}^{\dag}Y_{N}Y_{{\eta}}^{\dag} }
- 8 \tr{Y_{{\eta}}Y_{{\eta}}^{\dag}\T{Y_{N} }\conj{Y_{N}} }
- 8 \tr{Y_{N}Y_{{\eta}}^{\dag}Y_{{\eta}}Y_{N}^{\dag} } \notag \\
&- 8 \tr{\T{Y_{N} }\conj{Y_{N}}Y_{{\eta}}Y_{{\eta}}^{\dag} }, \notag \\
16 \pi^2 \beta_{\lambda_\phi} &= 18 \lambda_{\phi}^{2} + \frac{1}{2} \lambda_{{\phi\eta}}^{2} +2 \lambda_{{h\phi}}^{2} + 16 \lambda_{\phi} \tr{Y_{{\phi}}^{\dag}Y_{{\phi}} } - 64 \tr{Y_{{\phi}}Y_{{\phi}}^{\dag}Y_{{\phi}}Y_{{\phi}}^{\dag} }, \\
16 \pi^2 \beta_{\lambda_\eta} &= 18 \lambda_{\eta}^{2} + \frac{1}{2} \lambda_{{\phi\eta}}^{2} +2 \lambda_{{h\eta}}^{2} + 16 \lambda_{\eta} \tr{Y_{{\eta}}^{\dag}Y_{{\eta}} }
- 64 \tr{Y_{{\eta}}Y_{{\eta}}^{\dag}Y_{{\eta}}Y_{{\eta}}^{\dag} },\\
16 \pi^2 \beta_{\lambda_{\phi\eta}} &= 4 \lambda_{{\phi\eta}}^{2} + 4 \lambda_{{h\eta}} \lambda_{{h\phi}} + 6 \lambda_{\phi} \lambda_{{\phi\eta}} +6 \lambda_{\eta} \lambda_{{\phi\eta}} +8 \lambda_{{\phi\eta}} \tr{Y_{{\eta}}^{\dag}Y_{{\eta}} } + 8 \lambda_{{\phi\eta}} \tr{Y_{{\phi}}^{\dag}Y_{{\phi}} } \\
&- 64 \tr{Y_{{\eta}}Y_{{\eta}}^{\dag}Y_{{\phi}}Y_{{\phi}}^{\dag} } - 64 \tr{Y_{{\phi}}Y_{{\eta}}^{\dag}Y_{{\eta}}Y_{{\phi}}^{\dag} }
 - 64 \tr{Y_{{\eta}}Y_{{\phi}}^{\dag}Y_{{\phi}}Y_{{\eta}}^{\dag} } - 64 \tr{Y_{{\phi}}Y_{{\phi}}^{\dag}Y_{{\eta}}Y_{{\eta}}^{\dag} } \\
& - 64 \tr{Y_{{\phi}}Y_{{\eta}}^{\dag}Y_{{\phi}}Y_{{\eta}}^{\dag} }- 64 \tr{Y_{{\eta}}Y_{{\phi}}^{\dag}Y_{{\eta}}Y_{{\phi}}^{\dag} }, \notag \\
16 \pi^2 \beta_{Y_{N}} &= - \frac{3}{4} g^{2} Y_{N} - \frac{9}{4} g^{\prime 2} Y_{N} + \frac{3}{2} \tr{\left (Y_{{u}}^{\dag}Y_{{u}} \right )} Y_{N} + 2 Y_{N} Y_{{\phi}}^{\dag} Y_{{\phi}} +\frac{3}{2} Y_{N} Y_{N}^{\dag} Y_{N}
\\
& - \frac{3}{2} Y_{{e}}  Y_{{e}}^{\dag} Y_{N}+\frac{1}{2} \tr{\left (Y_{{e}}^{\dag}Y_{{e}} \right )} Y_{N} + \frac{3}{2} \tr{\left (Y_{{d}}^{\dag}Y_{{d}} \right )} Y_{N} + 2 Y_{N} Y_{{\eta}}^{\dag} Y_{{\eta}}  +\frac{3}{2} \tr{\left (\conj{Y_{{d}}}\T{Y_{{d}} } \right )} Y_{N} \notag \\
&+\frac{1}{2} \tr{\left (\conj{Y_{{e}}}\T{Y_{{e}}} \right )} Y_{N}  +\frac{3}{2} \tr{\left (\conj{Y_{{u}}}\T{Y_{{u}} } \right )} Y_{N}  \notag + \frac{1}{2} \tr{\left (Y_{N}^{\dag}Y_{N} \right )} Y_{N} +\frac{1}{2} \tr{\left (\conj{Y_{N}}\T{Y_{N} } \right )} Y_{N}, \notag \\
16 \pi^2 \beta_{Y_\phi} &= 4 \tr{\left (Y_{{\phi}}^{\dag}Y_{{\phi}} \right )} Y_{\phi}
+ 8 Y_{{\eta}} Y_{{\phi}}^{\dag} Y_{{\eta}} + 2 \tr{\left (Y_{{\eta}}^{\dag}Y_{{\phi}} \right )} Y_{\eta} + 2 \tr{\left (Y_{{\phi}}^{\dag}Y_{{\eta}} \right )} Y_{\eta}
 \\
& +\T{Y_{N}} \conj{Y_{N}} Y_{{\phi}} + 2 Y_{{\eta}} Y_{{\eta}}^{\dag} Y_{{\phi}} + Y_{{\phi}} Y_{N}^{\dag} Y_{N} +12 Y_{{\phi}} Y_{{\phi}}^{\dag}  Y_{{\phi}} +Y_{{\phi}} Y_{{\eta}}^{\dag} Y_{{\eta}}, \notag \\
16 \pi^2 \beta_{Y_\eta} &= 4 \tr{\left (Y_{{\eta}}^{\dag}Y_{{\eta}} \right )} Y_{\eta}
+ 8 Y_{{\phi}} Y_{{\eta}}^{\dag} Y_{{\phi}} + 2 \tr{\left (Y_{{\phi}}^{\dag}Y_{{\eta}} \right )} Y_{\phi} + 2 \tr{\left (Y_{{\eta}}^{\dag}Y_{{\phi}} \right )} Y_{\phi}
 \\
& +\T{Y_{N}} \conj{Y_{N}} Y_{{\eta}} + 2 Y_{{\phi}} Y_{{\phi}}^{\dag} Y_{{\eta}} + Y_{{\eta}} Y_{N}^{\dag} Y_{N}
 +12 Y_{{\eta}} Y_{{\eta}}^{\dag}  Y_{{\eta}} +Y_{{\eta}} Y_{{\phi}}^{\dag} Y_{{\phi}}, \notag \\
16 \pi^2 \beta_{Y_{e}} &= \beta_{Y_{e}}^{\rm SM}
+ \frac{1}{2} (\hc{Y_{N}} Y_{N} + \tr{ \conj{Y_{N}} \T{Y_{N}} }) Y_{e} - \frac{3}{2} Y_{N} \hc{Y_{N}} Y_{e}, \\
16 \pi^2 \beta_{Y_{d}} &= \beta_{Y_{d}}^{\rm SM} + \frac{1}{2} (\hc{Y_{N}} Y_{N} + \tr{ \conj{Y_{N}} \T{Y_{N}} }) Y_{d}, \\
16 \pi^2 \beta_{Y_{u}} &= \beta_{Y_{u}}^{\rm SM} + \frac{1}{2} (\hc{Y_{N}} Y_{N} + \tr{ \conj{Y_{N}} \T{Y_{N}} }) Y_{u}.
\end{align}

\bibliographystyle{JHEP}
\bibliography{citations}

\providecommand{\href}[2]{#2}\begingroup\raggedright\begin{thebibliography}{100}

\bibitem{Ade:2014xna}
{\bf BICEP2 Collaboration} Collaboration, P.~Ade et~al., {\it {BICEP2 I:
  Detection Of B-mode Polarization at Degree Angular Scales}},
  \href{http://xxx.lanl.gov/abs/1403.3985}{{\tt arXiv:1403.3985}}.

\bibitem{Guth:1980zm}
A.~H. Guth, {\it {The Inflationary Universe: A Possible Solution to the Horizon
  and Flatness Problems}},  {\em Phys.Rev.} {\bf D23} (1981) 347--356.

\bibitem{Linde:1981mu}
A.~D. Linde, {\it {A New Inflationary Universe Scenario: A Possible Solution of
  the Horizon, Flatness, Homogeneity, Isotropy and Primordial Monopole
  Problems}},  {\em Phys.Lett.} {\bf B108} (1982) 389--393.

\bibitem{Albrecht:1982wi}
A.~Albrecht and P.~J. Steinhardt, {\it {Cosmology for Grand Unified Theories
  with Radiatively Induced Symmetry Breaking}},  {\em Phys.Rev.Lett.} {\bf 48}
  (1982) 1220--1223.

\bibitem{Mukhanov:1981xt}
V.~F. Mukhanov and G.~V. Chibisov, {\it {Quantum Fluctuation and Nonsingular
  Universe. (In Russian)}},  {\em JETP Lett.} {\bf 33} (1981) 532--535.

\bibitem{Bardeen:1983qw}
J.~M. Bardeen, P.~J. Steinhardt, and M.~S. Turner, {\it {Spontaneous Creation
  of Almost Scale - Free Density Perturbations in an Inflationary Universe}},
  {\em Phys.Rev.} {\bf D28} (1983) 679.

\bibitem{Mukhanov:1990me}
V.~F. Mukhanov, H.~Feldman, and R.~H. Brandenberger, {\it {Theory of
  cosmological perturbations. Part 1. Classical perturbations. Part 2. Quantum
  theory of perturbations. Part 3. Extensions}},  {\em Phys.Rept.} {\bf 215}
  (1992) 203--333.

\bibitem{Olive:1989nu}
K.~A. Olive, {\it {Inflation}},  {\em Phys.Rept.} {\bf 190} (1990) 307--403.

\bibitem{Lyth:1998xn}
D.~H. Lyth and A.~Riotto, {\it {Particle physics models of inflation and the
  cosmological density perturbation}},  {\em Phys.Rept.} {\bf 314} (1999)
  1--146, [\href{http://xxx.lanl.gov/abs/hep-ph/9807278}{{\tt
  hep-ph/9807278}}].

\bibitem{2009arXiv0902.1529K}
W.~H. {Kinney}, {\it {TASI Lectures on Inflation}},  {\em ArXiv e-prints}
  (Feb., 2009) [\href{http://xxx.lanl.gov/abs/0902.1529}{{\tt
  arXiv:0902.1529}}].

\bibitem{Ade:2013uln}
{\bf Planck Collaboration} Collaboration, P.~Ade et~al., {\it {Planck 2013
  results. XXII. Constraints on inflation}},
  \href{http://xxx.lanl.gov/abs/1303.5082}{{\tt arXiv:1303.5082}}.

\bibitem{Mohanty:2014kwa}
S.~Mohanty and A.~Nautiyal, {\it {Signature of Gibbons-Hawking temperature in
  the BICEP2 measurement of gravitational waves}},
  \href{http://xxx.lanl.gov/abs/1404.2222}{{\tt arXiv:1404.2222}}.

\bibitem{Lyth:1996im}
D.~H. Lyth, {\it {What would we learn by detecting a gravitational wave signal
  in the cosmic microwave background anisotropy?}},  {\em Phys.Rev.Lett.} {\bf
  78} (1997) 1861--1863, [\href{http://xxx.lanl.gov/abs/hep-ph/9606387}{{\tt
  hep-ph/9606387}}].

\bibitem{Boubekeur:2005zm}
L.~Boubekeur and D.~Lyth, {\it {Hilltop inflation}},  {\em JCAP} {\bf 0507}
  (2005) 010, [\href{http://xxx.lanl.gov/abs/hep-ph/0502047}{{\tt
  hep-ph/0502047}}].

\bibitem{Okada:2014lxa}
N.~Okada, V.~N. {\c S}eno{\u g}uz, and Q.~Shafi, {\it {Simple Inflationary
  Models in Light of BICEP2: an Update}},
  \href{http://xxx.lanl.gov/abs/1403.6403}{{\tt arXiv:1403.6403}}.

\bibitem{Lyth:2014yya}
D.~H. Lyth, {\it {BICEP2, the curvature perturbation and supersymmetry}},
  \href{http://xxx.lanl.gov/abs/1403.7323}{{\tt arXiv:1403.7323}}.

\bibitem{Kehagias:2014wza}
A.~Kehagias and A.~Riotto, {\it {Remarks about the Tensor Mode Detection by the
  BICEP2 Collaboration and the Super-Planckian Excursions of the Inflaton
  Field}},  \href{http://xxx.lanl.gov/abs/1403.4811}{{\tt arXiv:1403.4811}}.

\bibitem{Choudhury:2013jya}
S.~Choudhury, A.~Mazumdar, and S.~Pal, {\it {Low \& High scale MSSM inflation,
  gravitational waves and constraints from Planck}},  {\em JCAP} {\bf 1307}
  (2013) 041, [\href{http://xxx.lanl.gov/abs/1305.6398}{{\tt
  arXiv:1305.6398}}].

\bibitem{Choudhury:2013iaa}
S.~Choudhury and A.~Mazumdar, {\it {An accurate bound on tensor-to-scalar ratio
  and the scale of inflation}},  {\em Nucl.Phys.} {\bf B882} (2014) 386--396,
  [\href{http://xxx.lanl.gov/abs/1306.4496}{{\tt arXiv:1306.4496}}].

\bibitem{Choudhury:2014kma}
S.~Choudhury and A.~Mazumdar, {\it {Reconstructing inflationary potential from
  BICEP2 and running of tensor modes}},
  \href{http://xxx.lanl.gov/abs/1403.5549}{{\tt arXiv:1403.5549}}.

\bibitem{Antusch:2014cpa}
S.~Antusch and D.~Nolde, {\it {BICEP2 implications for single-field slow-roll
  inflation revisited}},  \href{http://xxx.lanl.gov/abs/1404.1821}{{\tt
  arXiv:1404.1821}}.

\bibitem{Calmet:2014lga}
X.~Calmet and V.~Sanz, {\it {Excursion into Quantum Gravity via Inflation}},
  \href{http://xxx.lanl.gov/abs/1403.5100}{{\tt arXiv:1403.5100}}.

\bibitem{Branchina:2013jra}
V.~Branchina and E.~Messina, {\it {Stability, Higgs Boson Mass and New
  Physics}},  {\em Phys.Rev.Lett.} {\bf 111} (2013) 241801,
  [\href{http://xxx.lanl.gov/abs/1307.5193}{{\tt arXiv:1307.5193}}].

\bibitem{Bezrukov:2007ep}
F.~L. Bezrukov and M.~Shaposhnikov, {\it {The Standard Model Higgs boson as the
  inflaton}},  {\em Phys.Lett.} {\bf B659} (2008) 703--706,
  [\href{http://xxx.lanl.gov/abs/0710.3755}{{\tt arXiv:0710.3755}}].

\bibitem{Degrassi:2012ry}
G.~Degrassi, S.~Di~Vita, J.~Elias-Miro, J.~R. Espinosa, G.~F. Giudice, et~al.,
  {\it {Higgs mass and vacuum stability in the Standard Model at NNLO}},  {\em
  JHEP} {\bf 1208} (2012) 098, [\href{http://xxx.lanl.gov/abs/1205.6497}{{\tt
  arXiv:1205.6497}}].

\bibitem{Bezrukov:2012sa}
F.~Bezrukov, M.~Y. Kalmykov, B.~A. Kniehl, and M.~Shaposhnikov, {\it {Higgs
  Boson Mass and New Physics}},  {\em JHEP} {\bf 1210} (2012) 140,
  [\href{http://xxx.lanl.gov/abs/1205.2893}{{\tt arXiv:1205.2893}}].

\bibitem{Buttazzo:2013uya}
D.~Buttazzo, G.~Degrassi, P.~P. Giardino, G.~F. Giudice, F.~Sala, et~al., {\it
  {Investigating the near-criticality of the Higgs boson}},  {\em JHEP} {\bf
  1312} (2013) 089, [\href{http://xxx.lanl.gov/abs/1307.3536}{{\tt
  arXiv:1307.3536}}].

\bibitem{Kadastik:2011aa}
M.~Kadastik, K.~Kannike, A.~Racioppi, and M.~Raidal, {\it {Implications of the
  125 GeV Higgs boson for scalar dark matter and for the CMSSM phenomenology}},
   {\em JHEP} {\bf 1205} (2012) 061,
  [\href{http://xxx.lanl.gov/abs/1112.3647}{{\tt arXiv:1112.3647}}].

\bibitem{Lebedev:2012zw}
O.~Lebedev, {\it {On Stability of the Electroweak Vacuum and the Higgs
  Portal}},  {\em Eur.Phys.J.} {\bf C72} (2012) 2058,
  [\href{http://xxx.lanl.gov/abs/1203.0156}{{\tt arXiv:1203.0156}}].

\bibitem{EliasMiro:2012ay}
J.~Elias-Miro, J.~R. Espinosa, G.~F. Giudice, H.~M. Lee, and A.~Strumia, {\it
  {Stabilization of the Electroweak Vacuum by a Scalar Threshold Effect}},
  {\em JHEP} {\bf 1206} (2012) 031,
  [\href{http://xxx.lanl.gov/abs/1203.0237}{{\tt arXiv:1203.0237}}].

\bibitem{Gabrielli:2013hma}
E.~Gabrielli, M.~Heikinheimo, K.~Kannike, A.~Racioppi, M.~Raidal, et~al., {\it
  {Towards Completing the Standard Model: Vacuum Stability, EWSB and Dark
  Matter}},  {\em Phys.Rev.} {\bf D89} (2014) 015017,
  [\href{http://xxx.lanl.gov/abs/1309.6632}{{\tt arXiv:1309.6632}}].

\bibitem{Ko:2014eia}
P.~Ko and W.-I. Park, {\it {Higgs-portal assisted Higgs inflation in light of
  BICEP2}},  \href{http://xxx.lanl.gov/abs/1405.1635}{{\tt arXiv:1405.1635}}.

\bibitem{Froggatt:1995rt}
C.~Froggatt and H.~B. Nielsen, {\it {Standard model criticality prediction: Top
  mass 173 +- 5-GeV and Higgs mass 135 +- 9-GeV}},  {\em Phys.Lett.} {\bf B368}
  (1996) 96--102, [\href{http://xxx.lanl.gov/abs/hep-ph/9511371}{{\tt
  hep-ph/9511371}}].

\bibitem{Isidori:2007vm}
G.~Isidori, V.~S. Rychkov, A.~Strumia, and N.~Tetradis, {\it {Gravitational
  corrections to standard model vacuum decay}},  {\em Phys.Rev.} {\bf D77}
  (2008) 025034, [\href{http://xxx.lanl.gov/abs/0712.0242}{{\tt
  arXiv:0712.0242}}].

\bibitem{Kobakhidze:2014xda}
A.~Kobakhidze and A.~Spencer-Smith, {\it {The Higgs vacuum is unstable}},
  \href{http://xxx.lanl.gov/abs/1404.4709}{{\tt arXiv:1404.4709}}.

\bibitem{Spencer-Smith:2014woa}
A.~Spencer-Smith, {\it {Higgs Vacuum Stability in a Mass-Dependent
  Renormalisation Scheme}},  \href{http://xxx.lanl.gov/abs/1405.1975}{{\tt
  arXiv:1405.1975}}.

\bibitem{Haba:2014zda}
N.~Haba and R.~Takahashi, {\it {Higgs inflation with singlet scalar dark matter
  and right-handed neutrino in the light of BICEP2}},
  \href{http://xxx.lanl.gov/abs/1404.4737}{{\tt arXiv:1404.4737}}.

\bibitem{Masina:2014yga}
I.~Masina, {\it {The Gravitational Wave Background and Higgs False Vacuum
  Inflation}},  \href{http://xxx.lanl.gov/abs/1403.5244}{{\tt
  arXiv:1403.5244}}.

\bibitem{Cook:2014dga}
J.~L. Cook, L.~M. Krauss, A.~J. Long, and S.~Sabharwal, {\it {Is Higgs
  Inflation Dead?}},  \href{http://xxx.lanl.gov/abs/1403.4971}{{\tt
  arXiv:1403.4971}}.

\bibitem{Hamada:2014iga}
Y.~Hamada, H.~Kawai, K.-y. Oda, and S.~C. Park, {\it {Higgs inflation still
  alive}},  \href{http://xxx.lanl.gov/abs/1403.5043}{{\tt arXiv:1403.5043}}.

\bibitem{Nakayama:2014koa}
K.~Nakayama and F.~Takahashi, {\it {Higgs Chaotic Inflation and the Primordial
  B-mode Polarization Discovered by BICEP2}},
  \href{http://xxx.lanl.gov/abs/1403.4132}{{\tt arXiv:1403.4132}}.

\bibitem{Fairbairn:2014nxa}
M.~Fairbairn, P.~Grothaus, and R.~Hogan, {\it {The Problem with False Vacuum
  Higgs Inflation}},  \href{http://xxx.lanl.gov/abs/1403.7483}{{\tt
  arXiv:1403.7483}}.

\bibitem{Bezrukov:2014bra}
F.~Bezrukov and M.~Shaposhnikov, {\it {Higgs inflation at the critical point}},
   \href{http://xxx.lanl.gov/abs/1403.6078}{{\tt arXiv:1403.6078}}.

\bibitem{Enqvist:2014bua}
K.~Enqvist, T.~Meriniemi, and S.~Nurmi, {\it {Higgs Dynamics during
  Inflation}},  \href{http://xxx.lanl.gov/abs/1404.3699}{{\tt
  arXiv:1404.3699}}.

\bibitem{Salvio:2014soa}
A.~Salvio and A.~Strumia, {\it {Agravity}},
  \href{http://xxx.lanl.gov/abs/1403.4226}{{\tt arXiv:1403.4226}}.

\bibitem{Kaloper:2014zba}
N.~Kaloper and A.~Lawrence, {\it {Natural Chaotic Inflation and UV
  Sensitivity}},  \href{http://xxx.lanl.gov/abs/1404.2912}{{\tt
  arXiv:1404.2912}}.

\bibitem{Chialva:2014rla}
D.~Chialva and A.~Mazumdar, {\it {Super-Planckian excursions of the inflaton
  and quantum corrections}},  \href{http://xxx.lanl.gov/abs/1405.0513}{{\tt
  arXiv:1405.0513}}.

\bibitem{Linde:1982zj}
A.~D. Linde, {\it {Coleman-Weinberg Theory and a New Inflationary Universe
  Scenario}},  {\em Phys.Lett.} {\bf B114} (1982) 431.

\bibitem{Ellis:1982ws}
J.~R. Ellis, D.~V. Nanopoulos, K.~A. Olive, and K.~Tamvakis, {\it {PRIMORDIAL
  SUPERSYMMETRIC INFLATION}},  {\em Nucl.Phys.} {\bf B221} (1983) 524.

\bibitem{Ellis:1982dg}
J.~R. Ellis, D.~V. Nanopoulos, K.~A. Olive, and K.~Tamvakis, {\it {Fluctuations
  in a Supersymmetric Inflationary Universe}},  {\em Phys.Lett.} {\bf B120}
  (1983) 331.

\bibitem{Coleman:1973jx}
S.~R. Coleman and E.~J. Weinberg, {\it {Radiative Corrections as the Origin of
  Spontaneous Symmetry Breaking}},  {\em Phys.Rev.} {\bf D7} (1973) 1888--1910.

\bibitem{Langbein:1993ym}
R.~Langbein, K.~Langfeld, H.~Reinhardt, and L.~von Smekal, {\it {Natural slow
  roll inflation}},  {\em Mod.Phys.Lett.} {\bf A11} (1996) 631--646,
  [\href{http://xxx.lanl.gov/abs/hep-ph/9310335}{{\tt hep-ph/9310335}}].

\bibitem{GonzalezDiaz:1986bu}
P.~Gonzalez-Diaz, {\it {PRIMORDIAL KALUZA-KLEIN INFLATION}},  {\em Phys.Lett.}
  {\bf B176} (1986) 29--32.

\bibitem{Yokoyama:1998rw}
J.~Yokoyama, {\it {Chaotic new inflation and primordial spectrum of adiabatic
  fluctuations}},  {\em Phys.Rev.} {\bf D59} (1999) 107303.

\bibitem{Rehman:2008qs}
M.~U. Rehman, Q.~Shafi, and J.~R. Wickman, {\it {GUT Inflation and Proton Decay
  after WMAP5}},  {\em Phys.Rev.} {\bf D78} (2008) 123516,
  [\href{http://xxx.lanl.gov/abs/0810.3625}{{\tt arXiv:0810.3625}}].

\bibitem{Barenboim:2013wra}
G.~Barenboim, E.~J. Chun, and H.~M. Lee, {\it {Coleman-Weinberg Inflation in
  light of Planck}},  {\em Phys.Lett.} {\bf B730} (2014) 81--88,
  [\href{http://xxx.lanl.gov/abs/1309.1695}{{\tt arXiv:1309.1695}}].

\bibitem{Okada:2013vxa}
N.~Okada and Q.~Shafi, {\it {Observable Gravity Waves From $U(1)_{B-L}$ Higgs
  and Coleman-Weinberg Inflation}},
  \href{http://xxx.lanl.gov/abs/1311.0921}{{\tt arXiv:1311.0921}}.

\bibitem{Heikinheimo:2013fta}
M.~Heikinheimo, A.~Racioppi, M.~Raidal, C.~Spethmann, and K.~Tuominen, {\it
  {Physical Naturalness and Dynamical Breaking of Classical Scale Invariance}},
   \href{http://xxx.lanl.gov/abs/1304.7006}{{\tt arXiv:1304.7006}}.

\bibitem{Chatrchyan:2012ufa}
{\bf CMS} Collaboration, S.~Chatrchyan et~al., {\it {Observation of a new boson
  at a mass of 125 GeV with the CMS experiment at the LHC}},  {\em Phys.Lett.}
  {\bf B716} (2012) 30--61, [\href{http://xxx.lanl.gov/abs/1207.7235}{{\tt
  arXiv:1207.7235}}].

\bibitem{Aad:2012tfa}
{\bf ATLAS} Collaboration, G.~Aad et~al., {\it {Observation of a new particle
  in the search for the Standard Model Higgs boson with the ATLAS detector at
  the LHC}},  {\em Phys.Lett.} {\bf B716} (2012) 1--29,
  [\href{http://xxx.lanl.gov/abs/1207.7214}{{\tt arXiv:1207.7214}}].

\bibitem{Hempfling:1996ht}
R.~Hempfling, {\it {The Next-to-minimal Coleman-Weinberg model}},  {\em
  Phys.Lett.} {\bf B379} (1996) 153--158,
  [\href{http://xxx.lanl.gov/abs/hep-ph/9604278}{{\tt hep-ph/9604278}}].

\bibitem{Foot:2007as}
R.~Foot, A.~Kobakhidze, and R.~R. Volkas, {\it {Electroweak Higgs as a
  pseudo-Goldstone boson of broken scale invariance}},  {\em Phys.Lett.} {\bf
  B655} (2007) 156--161, [\href{http://xxx.lanl.gov/abs/0704.1165}{{\tt
  arXiv:0704.1165}}].

\bibitem{Foot:2007ay}
R.~Foot, A.~Kobakhidze, K.~McDonald, and R.~Volkas, {\it {Neutrino mass in
  radiatively-broken scale-invariant models}},  {\em Phys.Rev.} {\bf D76}
  (2007) 075014, [\href{http://xxx.lanl.gov/abs/0706.1829}{{\tt
  arXiv:0706.1829}}].

\bibitem{Foot:2007iy}
R.~Foot, A.~Kobakhidze, K.~L. McDonald, and R.~R. Volkas, {\it {A Solution to
  the hierarchy problem from an almost decoupled hidden sector within a
  classically scale invariant theory}},  {\em Phys.Rev.} {\bf D77} (2008)
  035006, [\href{http://xxx.lanl.gov/abs/0709.2750}{{\tt arXiv:0709.2750}}].

\bibitem{Chang:2007ki}
W.-F. Chang, J.~N. Ng, and J.~M. Wu, {\it {Shadow Higgs from a scale-invariant
  hidden U(1)(s) model}},  {\em Phys.Rev.} {\bf D75} (2007) 115016,
  [\href{http://xxx.lanl.gov/abs/hep-ph/0701254}{{\tt hep-ph/0701254}}].

\bibitem{Holthausen:2009uc}
M.~Holthausen, M.~Lindner, and M.~A. Schmidt, {\it {Radiative Symmetry Breaking
  of the Minimal Left-Right Symmetric Model}},  {\em Phys.Rev.} {\bf D82}
  (2010) 055002, [\href{http://xxx.lanl.gov/abs/0911.0710}{{\tt
  arXiv:0911.0710}}].

\bibitem{Iso:2009ss}
S.~Iso, N.~Okada, and Y.~Orikasa, {\it {Classically conformal $B^-$ L extended
  Standard Model}},  {\em Phys.Lett.} {\bf B676} (2009) 81--87,
  [\href{http://xxx.lanl.gov/abs/0902.4050}{{\tt arXiv:0902.4050}}].

\bibitem{Iso:2009nw}
S.~Iso, N.~Okada, and Y.~Orikasa, {\it {The minimal B-L model naturally
  realized at TeV scale}},  {\em Phys.Rev.} {\bf D80} (2009) 115007,
  [\href{http://xxx.lanl.gov/abs/0909.0128}{{\tt arXiv:0909.0128}}].

\bibitem{Foot:2010av}
R.~Foot, A.~Kobakhidze, and R.~R. Volkas, {\it {Stable mass hierarchies and
  dark matter from hidden sectors in the scale-invariant standard model}},
  {\em Phys.Rev.} {\bf D82} (2010) 035005,
  [\href{http://xxx.lanl.gov/abs/1006.0131}{{\tt arXiv:1006.0131}}].

\bibitem{Holthausen:2013ota}
M.~Holthausen, J.~Kubo, K.~S. Lim, and M.~Lindner, {\it {Electroweak and
  Conformal Symmetry Breaking by a Strongly Coupled Hidden Sector}},  {\em
  JHEP} {\bf 1312} (2013) 076, [\href{http://xxx.lanl.gov/abs/1310.4423}{{\tt
  arXiv:1310.4423}}].

\bibitem{Steele:2013fka}
T.~Steele, Z.-W. Wang, D.~Contreras, and R.~Mann, {\it {Can Radiative Symmetry
  Breaking Generate Viable Dark Matter Mass and Abundance in a Scalar Singlet
  Higgs Portal Extension of the Standard Model?}},
  \href{http://xxx.lanl.gov/abs/1310.1960}{{\tt arXiv:1310.1960}}.

\bibitem{Carone:2013wla}
C.~D. Carone and R.~Ramos, {\it {Classical scale-invariance, the electroweak
  scale and vector dark matter}},  {\em Phys.Rev.} {\bf D88} (2013), no.~5
  055020, [\href{http://xxx.lanl.gov/abs/1307.8428}{{\tt arXiv:1307.8428}}].

\bibitem{Dermisek:2013pta}
R.~Dermisek, T.~H. Jung, and H.~D. Kim, {\it {Coleman-Weinberg Higgs}},
  \href{http://xxx.lanl.gov/abs/1308.0891}{{\tt arXiv:1308.0891}}.

\bibitem{Hambye:2013dgv}
T.~Hambye and A.~Strumia, {\it {Dynamical generation of the weak and Dark
  Matter scale}},  {\em Phys.Rev.} {\bf D88} (2013) 055022,
  [\href{http://xxx.lanl.gov/abs/1306.2329}{{\tt arXiv:1306.2329}}].

\bibitem{Radovcic:2014rea}
B.~Radovcic and S.~Benic, {\it {Electroweak breaking and Dark Matter from the
  common scale}},  {\em Phys.Lett.} {\bf B732} (2014) 91--94,
  [\href{http://xxx.lanl.gov/abs/1401.8183}{{\tt arXiv:1401.8183}}].

\bibitem{Englert:2013gz}
C.~Englert, J.~Jaeckel, V.~Khoze, and M.~Spannowsky, {\it {Emergence of the
  Electroweak Scale through the Higgs Portal}},  {\em JHEP} {\bf 1304} (2013)
  060, [\href{http://xxx.lanl.gov/abs/1301.4224}{{\tt arXiv:1301.4224}}].

\bibitem{Khoze:2013uia}
V.~V. Khoze, {\it {Inflation and Dark Matter in the Higgs Portal of Classically
  Scale Invariant Standard Model}},  {\em JHEP} {\bf 1311} (2013) 215,
  [\href{http://xxx.lanl.gov/abs/1308.6338}{{\tt arXiv:1308.6338}}].

\bibitem{Khoze:2013oga}
V.~V. Khoze and G.~Ro, {\it {Leptogenesis and Neutrino Oscillations in the
  Classically Conformal Standard Model with the Higgs Portal}},  {\em JHEP}
  {\bf 1310} (2013) 075, [\href{http://xxx.lanl.gov/abs/1307.3764}{{\tt
  arXiv:1307.3764}}].

\bibitem{Chun:2013soa}
E.~J. Chun, S.~Jung, and H.~M. Lee, {\it {Radiative generation of the Higgs
  potential}},  {\em Phys.Lett.} {\bf B725} (2013) 158--163,
  [\href{http://xxx.lanl.gov/abs/1304.5815}{{\tt arXiv:1304.5815}}].

\bibitem{Hashimoto:2013hta}
M.~Hashimoto, S.~Iso, and Y.~Orikasa, {\it {Radiative symmetry breaking at the
  Fermi scale and flat potential at the Planck scale}},  {\em Phys.Rev.} {\bf
  D89} (2014) 016019, [\href{http://xxx.lanl.gov/abs/1310.4304}{{\tt
  arXiv:1310.4304}}].

\bibitem{Khoze:2014xha}
V.~V. Khoze, C.~McCabe, and G.~Ro, {\it {Higgs vacuum stability from the dark
  matter portal}},  \href{http://xxx.lanl.gov/abs/1403.4953}{{\tt
  arXiv:1403.4953}}.

\bibitem{Guo:2014bha}
J.~Guo and Z.~Kang, {\it {Higgs Naturalness and Dark Matter Stability by Scale
  Invariance}},  \href{http://xxx.lanl.gov/abs/1401.5609}{{\tt
  arXiv:1401.5609}}.

\bibitem{Farzinnia:2014xia}
A.~Farzinnia and J.~Ren, {\it {Higgs Partner Searches and Dark Matter
  Phenomenology in Classically Scale Invariant Higgs Sector}},
  \href{http://xxx.lanl.gov/abs/1405.0498}{{\tt arXiv:1405.0498}}.

\bibitem{Hashimoto:2014ela}
M.~Hashimoto, S.~Iso, and Y.~Orikasa, {\it {Radiative Symmetry Breaking from
  Flat Potential in various U(1)' models}},  {\em Phys.Rev.} {\bf D89} (2014)
  056010, [\href{http://xxx.lanl.gov/abs/1401.5944}{{\tt arXiv:1401.5944}}].

\bibitem{Kubo:2014ova}
J.~Kubo, K.~S. Lim, and M.~Lindner, {\it {Electroweak Symmetry Breaking by
  QCD}},  \href{http://xxx.lanl.gov/abs/1403.4262}{{\tt arXiv:1403.4262}}.

\bibitem{Kubo:2014ida}
J.~Kubo, K.~S. Lim, and M.~Lindner, {\it {Gamma-ray Line from Nambu-Goldstone
  Dark Matter in a Scale Invariant Extension of the Standard Model}},
  \href{http://xxx.lanl.gov/abs/1405.1052}{{\tt arXiv:1405.1052}}.

\bibitem{Salopek:1988qh}
D.~Salopek, J.~Bond, and J.~M. Bardeen, {\it {Designing Density Fluctuation
  Spectra in Inflation}},  {\em Phys.Rev.} {\bf D40} (1989) 1753.

\bibitem{Fakir:1990eg}
R.~Fakir and W.~Unruh, {\it {Improvement on cosmological chaotic inflation
  through nonminimal coupling}},  {\em Phys.Rev.} {\bf D41} (1990) 1783--1791.

\bibitem{Kaiser:1994vs}
D.~I. Kaiser, {\it {Primordial spectral indices from generalized Einstein
  theories}},  {\em Phys.Rev.} {\bf D52} (1995) 4295--4306,
  [\href{http://xxx.lanl.gov/abs/astro-ph/9408044}{{\tt astro-ph/9408044}}].

\bibitem{Komatsu:1999mt}
E.~Komatsu and T.~Futamase, {\it {Complete constraints on a nonminimally
  coupled chaotic inflationary scenario from the cosmic microwave background}},
   {\em Phys.Rev.} {\bf D59} (1999) 064029,
  [\href{http://xxx.lanl.gov/abs/astro-ph/9901127}{{\tt astro-ph/9901127}}].

\bibitem{Nozari:2007eq}
K.~Nozari and S.~D. Sadatian, {\it {Non-Minimal Inflation after WMAP3}},  {\em
  Mod.Phys.Lett.} {\bf A23} (2008) 2933--2945,
  [\href{http://xxx.lanl.gov/abs/0710.0058}{{\tt arXiv:0710.0058}}].

\bibitem{future}
K.~Kannike, A.~Racioppi, and M.~Raidal, {\it In preparation},  {\em In
  preparation} (2014).

\bibitem{Guth:2007ng}
A.~H. Guth, {\it {Eternal inflation and its implications}},  {\em J.Phys.} {\bf
  A40} (2007) 6811--6826, [\href{http://xxx.lanl.gov/abs/hep-th/0702178}{{\tt
  hep-th/0702178}}].

\bibitem{Minkowski:1977sc}
P.~Minkowski, {\it {$\mu \to e \gamma$ at a Rate of One Out of 1-Billion Muon
  Decays?}},  {\em Phys.Lett.} {\bf B67} (1977) 421.

\bibitem{Yanagida}
T.~Yanagida in {\em {Baryon Number of the Universe and Unified Theories}},
  (Tsukuba, Japan), 13-14 Feb 1979.

\bibitem{GellMann:1980vs}
M.~Gell-Mann, P.~Ramond, and R.~Slansky, {\it {Complex Spinors and Unified
  Theories}},  {\em Conf.Proc.} {\bf C790927} (1979) 315--321,
  [\href{http://xxx.lanl.gov/abs/1306.4669}{{\tt arXiv:1306.4669}}].

\bibitem{Glashow:1979nm}
S.~Glashow, {\it {The Future of Elementary Particle Physics}},  {\em NATO
  Adv.Study Inst.Ser.B Phys.} {\bf 59} (1980) 687.

\bibitem{Mohapatra:1979ia}
R.~N. Mohapatra and G.~Senjanovic, {\it {Neutrino Mass and Spontaneous Parity
  Violation}},  {\em Phys.Rev.Lett.} {\bf 44} (1980) 912.

\bibitem{Schechter:1980gr}
J.~Schechter and J.~Valle, {\it {Neutrino Masses in SU(2) x U(1) Theories}},
  {\em Phys.Rev.} {\bf D22} (1980) 2227.

\bibitem{Schechter:1981cv}
J.~Schechter and J.~Valle, {\it {Neutrino Decay and Spontaneous Violation of
  Lepton Number}},  {\em Phys.Rev.} {\bf D25} (1982) 774.

\bibitem{Fukugita:1986hr}
M.~Fukugita and T.~Yanagida, {\it {Baryogenesis Without Grand Unification}},
  {\em Phys.Lett.} {\bf B174} (1986) 45.

\bibitem{Lyonnet:2013dna}
F.~Lyonnet, I.~Schienbein, F.~Staub, and A.~Wingerter, {\it {PyR@TE:
  Renormalization Group Equations for General Gauge Theories}},  {\em
  Comput.Phys.Commun.} {\bf 185} (2014) 1130--1152,
  [\href{http://xxx.lanl.gov/abs/1309.7030}{{\tt arXiv:1309.7030}}].

\bibitem{Enqvist:2013kaa}
K.~Enqvist, T.~Meriniemi, and S.~Nurmi, {\it {Generation of the Higgs
  Condensate and Its Decay after Inflation}},  {\em JCAP} {\bf 1310} (2013)
  057, [\href{http://xxx.lanl.gov/abs/1306.4511}{{\tt arXiv:1306.4511}}].

\bibitem{DeSimone:2012qr}
A.~De~Simone and A.~Riotto, {\it {Cosmological Perturbations from the Standard
  Model Higgs}},  {\em JCAP} {\bf 1302} (2013) 014,
  [\href{http://xxx.lanl.gov/abs/1208.1344}{{\tt arXiv:1208.1344}}].

\bibitem{Lyth:2005qk}
D.~H. Lyth, {\it {Generating the curvature perturbation at the end of
  inflation}},  {\em JCAP} {\bf 0511} (2005) 006,
  [\href{http://xxx.lanl.gov/abs/astro-ph/0510443}{{\tt astro-ph/0510443}}].

\bibitem{Dvali:2003em}
G.~Dvali, A.~Gruzinov, and M.~Zaldarriaga, {\it {A new mechanism for generating
  density perturbations from inflation}},  {\em Phys.Rev.} {\bf D69} (2004)
  023505, [\href{http://xxx.lanl.gov/abs/astro-ph/0303591}{{\tt
  astro-ph/0303591}}].

\bibitem{Kofman:2003nx}
L.~Kofman, {\it {Probing string theory with modulated cosmological
  fluctuations}},  \href{http://xxx.lanl.gov/abs/astro-ph/0303614}{{\tt
  astro-ph/0303614}}.

\bibitem{Lebedev:2012sy}
O.~Lebedev and A.~Westphal, {\it {Metastable Electroweak Vacuum: Implications
  for Inflation}},  {\em Phys.Lett.} {\bf B719} (2013) 415--418,
  [\href{http://xxx.lanl.gov/abs/1210.6987}{{\tt arXiv:1210.6987}}].

\bibitem{Giudice:2003jh}
G.~Giudice, A.~Notari, M.~Raidal, A.~Riotto, and A.~Strumia, {\it {Towards a
  complete theory of thermal leptogenesis in the SM and MSSM}},  {\em
  Nucl.Phys.} {\bf B685} (2004) 89--149,
  [\href{http://xxx.lanl.gov/abs/hep-ph/0310123}{{\tt hep-ph/0310123}}].

\end{thebibliography}\endgroup

\end{document}